\documentstyle[12pt,a41,psfig]{article}

\begin{document}

\begin{center}
{\LARGE\bf Renormalons and Higher-Twist Contributions to
Structure Functions}

\vspace{1cm}
{\underline{M.~Maul}$^a$,  E.~Stein$^b$, 
L.~Mankiewicz $^c$, M.~Meyer-Hermann$^d$,  and  A.~Sch\"afer$^a$} \\
\vspace*{1cm}
 {\it  $^a$ Institut f\"ur Theoretische Physik, Universit\"at Regensburg,
Universit\"atsstr. 31, D-93053 Regensburg, Germany} \\
\vspace*{3mm}
 {\it  $^b$ INFN, Sezione di Torino, Via P.~Giuria 1, I-10125 Torino, Italy} \\ 
\vspace*{3mm}
 {\it  $^c$  
Institut f\"ur Theoretische Physik, TU-M\"unchen, 
D-85747 Garching, Germany} \\ 
\vspace*{3mm}
 {\it  $^d$  Institut f\"ur Theoretische Physik, J.~W.~Goethe
Universit\"at Frankfurt,
\\ 
Postfach~11~19~32, D-60054~Frankfurt am Main, Germany} \\ 
\vspace*{2cm}

\end{center}

\begin{abstract}
\noindent
We review the possibility to use the renormalons emerging in the
perturbation series of the twist-2 part of the nonsinglet structure functions
$F_L$, $F_2$, $F_3$, and $g_1$ to make approximate predictions for the
magnitude of the appertaining twist-4 corrections. 
\end{abstract}

\vspace*{2cm}

\section{Introduction}
The precision of deep inelastic scattering experiments nowadays allows
for the disentanglement of genuine twist-4 corrections to unpolarized
structure functions. 
While the twist-2 parton density has a simple probabilistic
interpretation in the parton model, genuine 
twist-4 corrections can be seen as multiple particle correlations
between quarks and gluons.
For example  the twist-4 correction to the Bjorken sum rule $f^{(2)}$
can be interpreted as the interaction of the induced 
color electric $E_A^\sigma$ and color magnetic fields 
$B_A^\sigma$ of a single 
quark and  the corresponding nucleon remnant \cite{SGMS95}:
\begin{equation}
\langle p S | - B_A^\sigma j_A^0 + (\vec  j_A \times \vec E_A)^\sigma
|p S \rangle = 2 m_N^2 f^{(2)} S^\sigma \quad,
\end{equation}
where $m_N$ is the nucleon mass, $p$ the momentum and $S$ the spin
of the incoming nucleon. Despite of the capability
to isolate at least for the unpolarized case the twist-4 contributions
to structure functions in experiments 
\cite{VM92}, on the theoretical side the description of 
higher-twist contributions is still not satisfactory. 
Even though the operator product expansion is possible for inclusive
quantities its technical applicability is restricted to the lowest
moments of the structure functions \cite{Shu82}. 
The best and physically most stringent approach 
to estimate the magnitude of the matrix elements of the twist-4
operators, the lattice gauge theory,
suffers from the problem of operator mixing which still has to be
solved.
In this approach up to now higher-twist corrections were
calculated only in cases where such a mixing is forbidden by some quantum
numbers, e.g.~as in the case of the twist-3 correction to the Bjorken sum rule
\cite{Me96}.
In total, we are far away from getting hold of the twist-4 contribution 
to all moments and consequently from a description which gives 
the $x$-dependent contribution of the twist-4 correction to the whole structure
function.
\newline
\newline
The renormalon approach cannot really cure this problem because
of its approximative and partially very speculative nature, but it
offers a well defined model. This model was applied to the 
nonsinglet structure functions either in terms of the massive gluon
scheme or the running coupling scheme which in general 
have an one to one correspondence \cite{PBal95}.
Empirically, it has proven to be quite successful to reproduce at least in the 
large-$x$ range the  $x$ dependence of the twist-4 correction to the 
unpolarized
structure functions \cite{SMMS,DW96,MSSM}.
The success of this model is nowadays referred to as the `ultraviolet
dominance' of twist-4 operators \cite{BBM97}. 
\newline
\newline
In the following first two sections we will shortly outline
the definition and the basic features of the renormalon in QCD and
the technique of naive nonabelianization (NNA), which allows to reduce
the calculational effort necessary to the mere summation of vacuum
polarization bubbles. We also comment on  the very
important question of the scheme dependence of the renormalon and
in the last section we want to discuss the renormalon
contribution to the structure functions $g_1$, $F_L$, $F_2$, and $F_3$ in
detail and compare it to the measured data.  
For a review and references to the classical papers on renormalons
see \cite{classical}.
\section{What are Renormalons?}
\vspace{1mm}
\noindent
The operator product expansion predicts that a moment of an unpolarized
structure function $F$ ($F = F_2,F_L,F_3$) admits up to ${\cal O}(1/Q^4)$ the
following general form:
\begin{eqnarray}
M_n(F) &:=& \int_0^1 dx x^n F(x)  
\nonumber \\
&=& C_n  a_n^{twist=2}
+\frac{1}{Q^2} \left[M^2 C_n'  b_n^{twist=2}
+ C^{(1)}_n  b_{1,n}^{twist=4}+C^{(2)}_n  b_{2,n}^{twist=4}+..\right] \;. 
\end{eqnarray}
Each Wilson Coefficient (e.g.~$C_n$) can be written as an asymptotic
series in the strong coupling constant. 
\begin{eqnarray}
C_n = C_n(Q^2/\mu^2, a_s) = \sum_{m=0}^{m_0-1} C_{m,n} a_s^m + \Delta R;
\; a_s = 
\frac{\alpha_s}{4\pi} \;.
\label{minimal}
\end{eqnarray}
$m_0$ corresponds to the minimal term of the series and 
is given by the condition
\begin{eqnarray}
\label{minimum}
\left| \frac{ a_s C_{n,m_0}}{C_{n,m_0-1}} \right|>1\; \Rightarrow \Delta R = 
C_{n,m_0} a_s^{m_0}\;.
\end{eqnarray}
In a certain approximation the Wilson coefficient can be evaluated to 
all orders, i.e.~when we take the numbers of fermions to infinity, $N_F \to \infty$. 
One then can derive a closed formula for $C_{m,n}$
for all $m$ and $n$. Only the flavor nonsinglet case has been 
considered so far.
The Borel-transformed series
\begin{eqnarray}
B[C_n](s) = \sum_m C_{n,m}\frac{1}{m!} \left(\frac{s}{\beta_0}\right)^m \;,
\end{eqnarray}
with $\beta_0 = -\frac{1}{4\pi} \frac{2}{3}N_F$ $(N_F \to \infty)$, 
has the pole structure: 
\begin{eqnarray}
B[C_n](s) &=& \left(\frac{\mu^2 e^{-C}}{Q^2} \right)^s
\left( ...+ \frac{p_n^{IR_2}}{2-s} + \frac{p_n^{IR_1}}{1-s}    
+ \frac{p_n^{UV_1}}{1+s} + \frac{p_n^{UV_2}}{2+s}+ ... \right) \;,   
\end{eqnarray}
reflecting the fact that the QCD-perturbation series is not Borel
summable. 
The $p_n^{UV_j}$ and $p_n^{IR_j}$ define residua of  ultraviolet- and
infrared-renormalon poles, respectively. The existence of
infrared-renormalon poles makes the unambiguous
reconstruction of the summed series from its Borel representation impossible.
$C$ is the finite part of the fermion loop insertion into the gluon
propagator equal to $-5/3$ in the $\rm \overline{MS}$ scheme.
%
%
%
%
%
The Borel representation $B[C_n]$ can be used as the generating
function for the fixed order coefficients
\begin{eqnarray}
C_{n,m} = \beta_0^m \frac{d^m}{ds^m} B[C_n](s) \Bigg|_{s=0}\;,
\end{eqnarray}
which implies for large $m$ the following asymptotic behavior of the
coefficient functions

\begin{eqnarray}
C_{n,m}^{IR_1} & \sim &
p_n^{IR_1} \left( \frac{\mu^2 e^{-C}}{Q^2} \right) m! \beta_0^m
\nonumber \\
C_{n,m}^{IR_2} & \sim &
p_n^{IR_2} \left( \frac{\mu^2 e^{-C}}{Q^2} \right)^2 m! \beta_0^m 
\left(\frac{1}{2} \right)^m
\nonumber \\
C_{n,m}^{UV_1} & \sim &
p_n^{UV_1} \left( \frac{\mu^2 e^{-C}}{Q^2} \right)^{-1} m! \beta_0^m
(-)^m
\nonumber \\
C_{n,m}^{UV_2} & \sim &
p_n^{UV_2} \left( \frac{\mu^2 e^{-C}}{Q^2} \right)^{-2} m! \beta_0^m
\left(-\frac{1}{2} \right)^m \;.
\end{eqnarray}
The infrared renormalon pole nearest to the origin of the Borel
plane dominates the asymptotic expansion.
Its name originates from the fact
that in asymptotically free theories the origin of the factorial divergence of the series
can be traced to the integration over the 
low momentum region of Feynmann diagrams:
\begin{eqnarray}
\sum_{m=0}^\infty  \alpha_s(Q^2)^{m+1} m! 
\frac{1}{j} \left(\frac{\beta_0}{j} \right)^m 
=  
\int_0^{Q^2} \frac{d k^2}{k^2} \left( \frac{k^2}{Q^2} 
\right)^j \alpha_s(k^2)\;,
\end{eqnarray}
with
\begin{equation}
\alpha(k^2) = \frac{\alpha(Q^2)}{1+ \beta_0 \alpha(Q^2) \ln(k^2/Q^2)}\quad.
\end{equation}
The uncertainty of the asymptotic perturbation series 
can be either estimated by calculating the minimal term in the
expansion (\ref{minimal}) or by  taking  the imaginary part (divided
by $\pi$) of $B[C_n](s)$ \cite{BraunMoriond}:
\begin{eqnarray}
\Delta R &=& \frac{1}{\pi} {\cal I} \frac{1}{\beta_0} 
\int_0^\infty ds \exp \left(\frac{-s}{\beta_0 \alpha_s}\right) B[R](s)
\nonumber \\
&\sim& \pm p_n^{IR_1} \frac{1}{\beta_0}\left(\frac{\Lambda_C^2 e^{-C}}{Q^2}
\right) \;.
\end{eqnarray}
The undetermined sign is due to the two possible contour integrations,
below or above the pole. 

In a physical quantity the infrared-renormalon ambiguities have to cancel,
against similar ambiguities in the definition of twist-4 operators such that in
principle only the sum of both is well defined.  Operators of higher-twist and
therefore higher dimension may exhibit power-like UV divergences, and the
corresponding ambiguity can be extracted as the quadratic UV divergence of the
twist-4 operator. Because of this cancellation, the IR renormalon is in a
one-to-one correspondence with the quadratic UV divergence of the twist-4
operator. Taking the IR renormalon contibution as an estimate of real twist-4
matrix elements is therefore equivalent to the assumption that the latter are
dominated by their UV divergent part.

In practice, the all-orders calculation can be performed only in the 
large-$N_F$ limit, which corresponds to the QED case, where it produces an
ultraviolet renormalon. To make the connection to QCD one uses the 'Naive Non
Abelianization' (NNA) recipe \cite{NNA} and substitutes for the QED one-loop
$\beta$-function the corresponding QCD expression.
\begin{eqnarray}
\beta_0 ^{\rm limit} &=& -\frac{1}{4\pi} \frac{2}{3} N_F \rightarrow
\frac{1}{4\pi}\left( 11 -  \frac{2}{3} N_F  \right) \;.
\end{eqnarray}
The positive sign of $\beta_0$ in QCD produces an infrared-renormalon. 
While no satisfactory justification for this
approximation is known so far, it has proven to work quite well in
low orders were comparison to the known exact coefficients in the 
${\rm \overline{MS}}$ scheme is possible \cite{MMS}.
Despite its phenomenological success in describing the shape of higher-twist
corrections to various QCD observables, one should be aware of conceptual
limitations of this approach. As the renormalon contribution is constructed
from twist-2 parton distributions only, it really has no sensitivity to the
intrinsic non-perturbative twist-4 nucleon structure. For example the ratio of
the n-th moment $M_n$ of the twist-4 renormalon prediction to a structure
function and the n-th moment of the twist-2 part of the structure functions
itself is by construction the same for different hadrons:
\begin{eqnarray}
\frac{M_n^{twist-4}}{M_n^{twist-2}}\Bigg|_{\rm hadron 1}
-
\frac{M_n^{twist-4}}{M_n^{twist-2}}\Bigg|_{\rm hadron 2} = 0\;,
\end{eqnarray}
which is definitely not necessarily the case in reality.

As far as the scheme dependence of the renormalon model is concerned, its
prediction for the magnitude of the mass scale of higher-twist corrections
$\sim \Lambda_C^2 e^{-C}$ is scheme invariant in the large-$N_F$ limit only. On
the other hand, once one takes this scale to be a fit parameter, where 
the magnitude of this fit parameter
depends on the process under consideration, this scheme dependence
disappears. Even in this case, however, the mass scale still depends on the
scheme used to determine perturbative corrections to the twist-2 part.  Thus
one can only say that the renormalon model is just a very economical way of
estimating higher-twist contributions in the situation when their exact theory
has not yet been constructed.  In addition, in cases where the operator product
expansion is not directly applicable, like in Drell Yan, the renormalon can
still trace power corrections and give an approximation of their magnitude.  

In order to demonstrate how it works, let us consider the phenomenologically
important case of twist-4 corrections to the Bjorken sum-rule:
\begin{eqnarray}
\int_0^1 dx g_1^{p-n}(x,Q^2) &=& \frac{1}{6}
\frac{g_A}{g_V} + \mu \frac{GeV^2}{Q^2} + 
{\cal O}(1/Q^4)
\nonumber \\
g_A/g_V &=& -1.2601 \pm 0.0025 \qquad \cite{PPBL96}\;.
\end{eqnarray}
The $1/Q^2$ power corrections consist of target-mass corrections, a twist-3 and
a twist-4 part \cite{Bruno}. Only the twist-4 part can be traced by the
renormalon ambiguity because this is the only operator which can mix with
the spin-1, twist-2 axial current operator.  Table 1 contains results of
various approaches employed so far. A QCD-sum rule calculation is described in
\cite{BBK,SGMS95}.  Due to the high dimension of the corresponding operator the
error arising from the sum rule calculation is large, typically of the order of
50
percent.  Another possibility is the MIT-Bag model, which however lacks
explicit Lorentz invariance. As long as an unambiguous field-theoretical
definition of twist-4 matrix elements does not exist, only the 
twist-3 part of the
coefficient $\mu$ can be calculated on the lattice \cite{Me96}. 
Recently the twist-4 matrix element has been estimated within the 
framework of the chiral soliton model \cite{BPW97}, providing  
a semi classical description. 
The prediction
of the renormalon model, which attributes the whole power correction to the
twist-4 operator, is in an order-of-magnitude agreement with results of the
other methods.
The twist-4 contribution has also
been estimated from the experimental data \cite{JM97}, yielding a value
within a large error which is also in agreement with all the other
methods as far as the order of magnitude is concerned, but predicting
a positive sign in contrast to all calculations except the
bag-model calculation. Note, that due to its nature as an ambiguity
in the twist-2 calculation the renormalon prediction is not capable
to fix the absolute sign.
\newline
\newline
Having the calculational efforts of sum rules and lattice QCD in
mind, it might be an interesting idea to get a zero-order guess of
the size of the twist-4 corrections by looking at the ambiguity in 
the expansion of the twist-2 part. 
\newline 
\begin{table}
\begin{center}
\begin{tabular}{|c|c|c|}
\hline
\hline
&&\\
method & value & reference \\
&&\\
\hline
\hline
&&\\
sum rule & $\mu=0.003\pm 0.008$ & \cite{SGMS95}\\
         & $\mu_{twist-4}= -0.0094 $ &\cite{SGMS95}\\
&&\\
bag model &  $\mu = 0.027$ & \cite{JU94} \\
         & $\mu_{twist-4}= 0.014$ &\cite{JU94}\\

&& \\
chiral soliton model &  $\mu_{twist-4}= -0.013$ &\cite{BPW97}\\
&&\\
extraction from data &  $\mu_{twist-4}=  0.012\pm 0.037$ &\cite{JM97}\\
&&\\
\underline{renormalon} & $\mu = \pm 0.017$ & \cite{Bra95} \\
&& \\
\hline
\hline
\end{tabular}
\end{center}
\caption{Higher-twist contributions to the
Bjorken sum rule} 
\end{table}
\section{The renormalon 
contribution to $F_L$,$F_2$,$F_3$, and
$g_1$}
\begin{figure}
\centerline{\psfig{figure=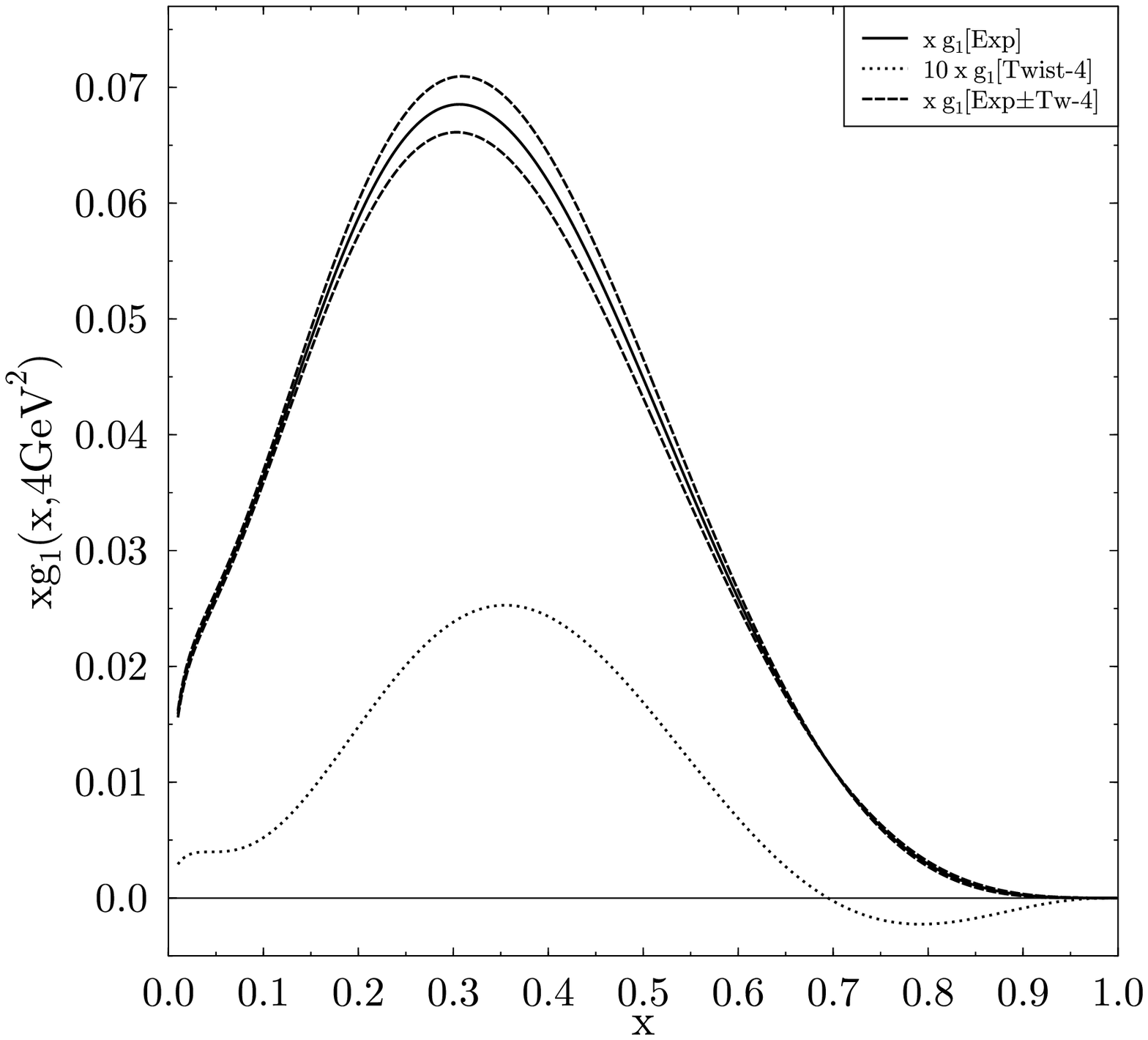,width=13cm}}
\caption{\sf
Experimental fit \cite{Geh95} for $g_1(x,4\,{\rm GeV}^2)$ (full line).
The estimate
for the twist-4 correction is given explicitly multiplied by a
factor of $10$ (dotted line). 
The correction was
added and subtracted from the
experimental values (dashed lines)
($\Lambda_{\overline{MS}}=200~{\rm MeV}$, 
$Q^2=4~{\rm GeV}$ and $N_f = 4$).}
\label{fig1}
\end{figure}

\begin{figure}
\centerline{\psfig{figure=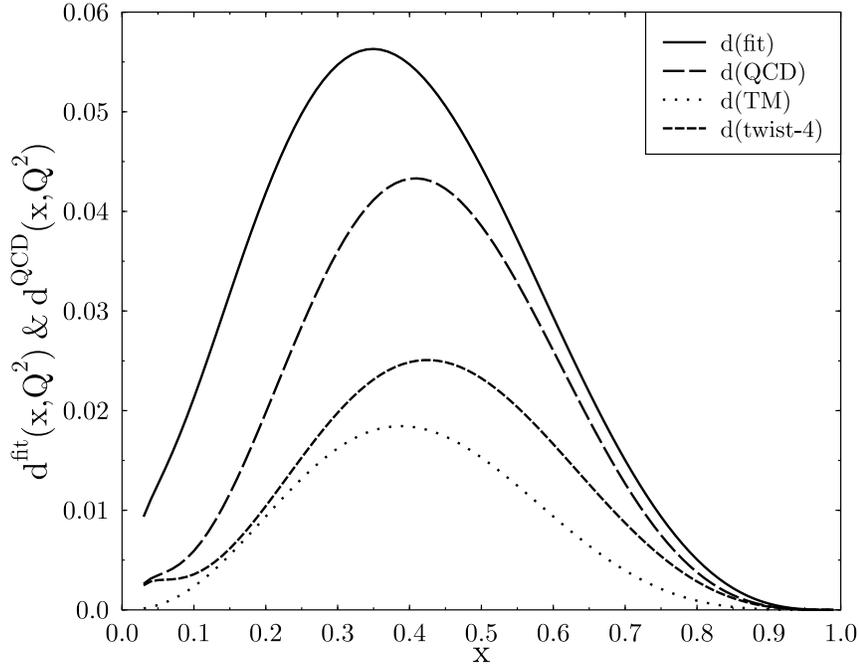,width=12cm}}
\caption
{$F_L (x,Q^2) = F_L^{twist-2}(x,Q^2) + \frac{d(x,Q^2)}{Q^2}$
$+ {\cal O}(1/Q^4)$. We have chosen
$\Lambda_{\overline{MS}}=250~{\rm MeV}$, $Q^2=5~{\rm GeV}$ and $N_f = 4$.}
\label{fig2}
\end{figure}

\begin{figure}
\centerline{\psfig{figure=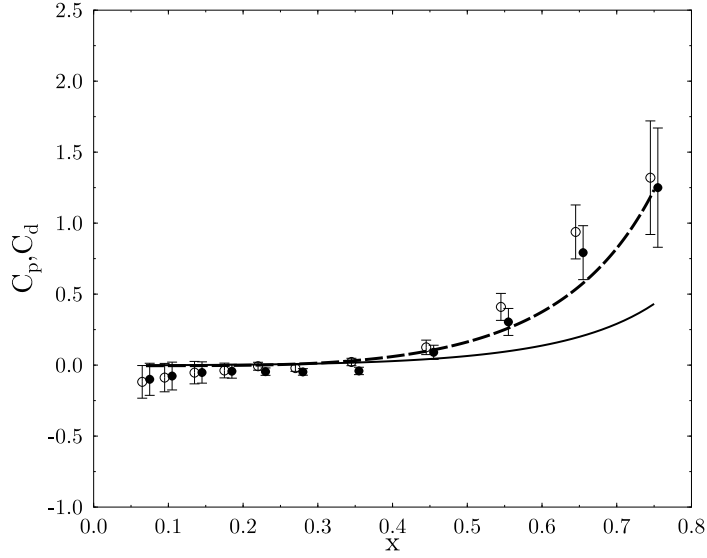,width=10cm}}
\caption{  
  The solid line shows the renormalon model predictions
  for $C_p$ and $C_d$, which are indistinguishable, LO parameterization taken
  from \cite{GRV94}. The dashed line shows the fit, as in
  Ref.~\cite{DW96}. The filled and empty circles display the
  data for $C_p$ and $C_d$ according to Ref.~\cite{VM92}, respectively. }
\label{fig3}
\end{figure}

\begin{figure}
\centerline{\psfig{figure=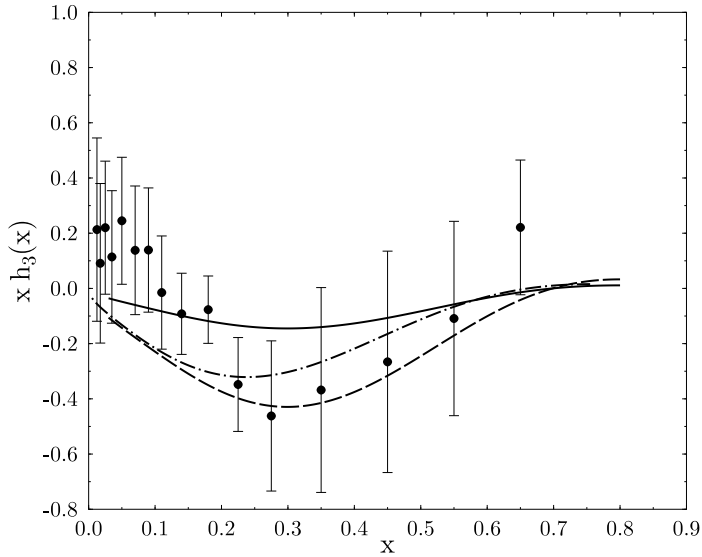,width=10cm}}
\caption{
  Renormalon prediction for $x h_3(x)$ using the LO GRV \cite{GRV94} 
  parametrization (solid line). The data points taken from Ref.~\cite{Si97}
  correspond to the NLO analysis, which provides the best fit of the 
  higher-twist corrections. However, to be consitent within the
  renormalon approach the parton distribution sets used for the renormalon
  prediction have to be LO. The dashed line shows the prediction with the
  scale $\mu^2$ adjusted to the description of the coefficients $C_p$ and
  $C_d$, as in \cite{DW96}. The dot-dashed line shows the original prediction
  of \cite{DW96}, obtained using the MRSA parametrization \cite{MRSA}
  normalized at $Q^2 = 10$ GeV$^2$.}
\label{fig4}
\end{figure}

Finally we will discuss the application of the renormalon model to the
$x$ dependence of DIS structure functions.
Indeed the real success of the renormalon method in DIS - as noticed in 
\cite{SMMS,DW96} - is  their
capability to reproduce the $x$ dependence of measured higher-twist 
contributions. The absolute magnitude of the renormalon contribution
can be either left as a fit parameter, taken from the NNA-calculation,
or taken as an universal constant as done in the gluon scheme of 
\cite{DMW96}.
\newline
\newline
We once more regard the nonsinglet part
of the structure function $g_1$ \cite{MMSMS}. In Fig.~\ref{fig1} $xg_1$ is 
plotted from the parton distribution set Gehrman/Stirling Gluon A \cite{Geh95}
plus/minus the renormalon predicted twist-4 contribution, which is on the  
percent level. The dotted line gives the renormalon prediction again,
amplified by a factor of 10. It becomes visible that 
the curve contains a zero at $x\approx0.7$, a feature which is a very
definite and clear prediction and should be confronted to experimental
data, if a precision on the precent level was reached.
\newline
\newline
As to $F_L$ \cite{SMMS}, we plot in Fig.~\ref{fig2} 
the appertaining renormalon prediction to the twist-4 operator 
(short dashed line). The target mass corrections taken
from \cite{Spanier} are plotted by a dotted line.
The sum of both is combined to the long dashed
line and has to be compared to the experminental fit by \cite{Whitlow,
NMC} (solid line).
The $x$ dependence is quite well approximated by the renormalon
approach, which predicts well the measured $x$ dependence 
although the absolute magnitude is smaller than 
what is suggested by the experimental fit,
but the $x$ dependence is described in an acceptable way.
\newline
\newline
As to the notation for the twist-4 corrections of the structure functions
$F_2$ and $F_3$ we write ($i = 2,3$)
\begin{eqnarray}
F_i( x,Q^2) &=& F_i^{(LT)}(x,Q^2)+  \frac{1}{Q^2} h_i(x) 
= F_i^{(LT)}(x,Q^2) \left( 1 + \frac{C_i(x)}{Q^2} \right) \;.
\end{eqnarray}
In the case of $F_2$ the good agreement between the  $x$ behavior of the 
nonsinglet renormalon contribution and the deuteron and 
proton twist-4 contribution
(see Fig~\ref{fig3}) to $F_2$ has been 
noticed in \cite{DW96}. It is remarkable that the uncalculated pure
singlet part seems to be small or proportional to the nonsinglet part.
At least in the large-$x$ range, where gluons contribute only a minor part, this
is understandable.
If we calculate the absolute value of the renormalon contributions
for $F_2$ (solid line) it falls short by a factor 2 or 3, compared with what
seems to be required by the data \cite{MSSM}. 
\newline
\newline
The same behavior is found for the structure function $F_3$, where
the twist-4 contribution is shown in Fig.~\ref{fig4}. Again, if one
is adjusting the absolute size of the renormalon contribution in
the same way as it was done for $F_2$, the result is in the large-$x$
range in a good agreement with the data. Even more remarkable is
the fact that the data indicate a change in sign in the large-$x$ region 
close to the one of the renormalon prediction.
\section{Conclusions}
The renormalon prescription provides a satisfactory
description of the  shape of the
measured values of the twist-4 contributions at large $x$ 
in the cases studied so far, leaving the absolute normalization
as a fit parameter. 
Empirically, it falls short
by a factor 2 - 3 as to the measured magnitude of the finite twist-4 
contributions, when it is calculated in the $\rm \overline{MS}$ scheme.

\newpage


\end{document}